\documentclass[a4paper,11pt]{article}
\usepackage{amssymb}
\usepackage{pos}
\usepackage{longtable}
\usepackage{booktabs,colortbl}
\usepackage{multirow}
\usepackage{amsmath}
\usepackage{bbm}
\usepackage[utf8]{inputenc}
\usepackage{pdflscape}
\usepackage{xspace}
\usepackage[caption=false]{subfig}
\usepackage{nicefrac}
\usepackage{graphicx}
\usepackage{mathtools}
\usepackage{nccfoots}
\usepackage{tikz}
\usetikzlibrary{calc,positioning}
\usepackage{bm}
\newlength{\bibitemsep}
\newlength{\bibparskip}
\let\oldthebibliography\thebibliography
\renewcommand\thebibliography[1]{%
  \oldthebibliography{#1}%
  \setlength{\parskip}{\bibitemsep}%
  \setlength{\itemsep}{\bibparskip}%
}

\usepackage[all,cmtip]{xy}
\newlength{\xywd}
\newcommand{\xyrightarrow}[2][]{%
  \sbox{0}{$\scriptstyle#1$}%
  \xywd=\wd0
  \sbox{0}{$\scriptstyle#2$}%
  \ifdim\wd0>\xywd \xywd=\wd0 \fi
  \xymatrix@C\dimexpr\xywd+1em\relax{{}\ar[r]^{#2}_{#1}&{}}%
}

\newcommand{\rep}[1]{\ensuremath\boldsymbol{#1}}
\newcommand{\crep}[1]{\ensuremath\bar{\boldsymbol{#1}}}
\newcommand{\Z}[1]{\ensuremath{\mathbbm{Z}_{#1}}} % z_N ->\Z{N}

\newcommand{\SL}[1]{\ensuremath{\mathrm{SL}(#1)}}

\newcommand{\I}{\mathrm{i}}

\newcommand{\CP}{\ensuremath{\mathcal{CP}}\xspace}

\newcommand{\vev}[1]{\ensuremath{\langle{#1}\rangle}}

\usepackage{xcolor}
\definecolor{darkgreen}{HTML}{109930}
\definecolor{pink}{rgb}{0.858, 0.188, 0.478}

\title{Anatomy of a top-down approach to discrete and modular flavor symmetry}
\author*[a]{Andreas Trautner}
\affiliation[a]{Max-Planck-Institut f{\"u}r Kernphysik \\ Saupfercheckweg 1, 69117 Heidelberg, Germany}

\emailAdd{trautner@mpi-hd.mpg.de}
\abstract{%
The framework of compactified heterotic string theory offers consistent ultraviolet (UV) completions of the Standard Model (SM) of particle physics. 
In this approach, the existence of flavor symmetries beyond the SM is imperative and the flavor symmetries can be derived 
from the top-down. Such a derivation uncovers a unified origin of traditional discrete flavor symmetries, discrete modular flavor symmetries, 
discrete R symmetries of supersymmetry, as well as CP symmetry -- altogether called the eclectic flavor symmetry. 
I will show a specific example of such a top-down derived eclectic flavor symmetry, discuss the different sources of breaking of the eclectic flavor symmetry, 
as well as the possible lessons for bottom-up flavor model building.}

\FullConference{%
  7th Symposium on Prospects in the Physics of Discrete Symmetries (DISCRETE 2020-2021)\\
  29th November - 3rd December 2021\\
 Bergen, Norway}

\begin{document}
\maketitle

\section{Introduction}
Embracing grand unification, a solution to the electroweak hierarchy problem as well as a consistent quantum theory of gravity, 
there is strong motivation from the bottom-up (BU) to consider string theory as UV completion of the SM. 
At the same time, it is crucial to ensure that the SM can indeed be consistently incorporated in a concrete realization of string theory and expose
the constraints and predictions that may arise from such a derivation.
A particularly advanced setup is provided by orbifold compactifications of the heterotic string \cite{Ibanez:1986tp,Ibanez:1987sn,Gross:1984dd,Gross:1985fr,Dixon:1985jw,Dixon:1986jc} 
which cannot only consistently host the (supersymmetric) SM~\cite{Buchmuller:2005jr} but may also shed light on one of its
most pressing puzzles by automatically including family repetition and flavor symmetries~\cite{Kobayashi:2006wq,Olguin-Trejo:2018wpw}.
Since a ``theory of everything'' has to be, in particular, a theory of flavor, there is a strong motivation
to understand the flavor puzzle of the SM from such a top-down (TD) perspective.
The purpose of this talk is to present new progress that has been achieved in consistently deriving 
the complete flavor symmetry from concrete string theory models and understand the different 
sources that contribute to its breaking in the infrared (IR).

\section{Types of discrete flavor symmetries and the eclectic symmetry}
The action of the 4D effective $\mathcal{N}=1$ SUSY theory
can schematically be written as (here, $K$: K\"ahler potential, $W$: Superpotential, $x$: spacetime, $\theta$: superspace, $\Phi:$ superfields, $T$: modulus)
\begin{equation}
\mathcal{S}=\int d^4x\,d^2\theta\,d^2\bar{\theta}\,K(T,\bar{T},\Phi,\bar{\Phi})+ \int d^4x\,d^2\theta\,W(T,\Phi)+ \int d^4x\,d^2\bar{\theta}\,\bar{W}(\bar{T},\bar{\Phi})\;.
\end{equation}
There are four categories of possible symmetries that differ by their effect on fields and coordinates:
\begin{itemize}
 \item ``Traditional'' flavor symmetries ``$G_\mathrm{traditional}$'', see e.g.~\cite{Ishimori:2010au}: $\Phi\mapsto \rho(\mathsf{g})\Phi$\,,\quad$\mathsf{g}\in G_\mathrm{traditional}$.
  \item Modular flavor symmetries ``$G_\mathrm{modular}$'' \cite{Feruglio:2017spp}: (partly cancel between $K$ and $W$)   
 \begin{equation}
\gamma:=\begin{pmatrix}a&b\\c&d\end{pmatrix}\in\mathrm{SL}(2,\Z{})\;,\quad\Phi\xmapsto{\gamma}(c\,T+d)^n \rho(\gamma)\Phi\;,\quad  T\xmapsto{\gamma}\frac{a\,T+b}{c\,T+d}\;.  
 \end{equation}
 In this case couplings are promoted to modular forms: $Y=Y(T)$, $Y(\gamma T)=\left(c\,T+d\right)^{k_Y}\rho_Y(\gamma)\,Y(T)$.
 \item R flavor symmetries ``$G_R$'' that differ for fields and their superpartners~\cite{Chen:2013dpa}. (cancel between $W$ and $d^2\theta$)
 \item General symmetries of the ``\CP'' type~\cite{Baur:2019kwi,Novichkov:2019sqv}: (partly cancel between $K$ and $W$ and $d^4x$)
 \begin{equation}
  \det\left[\,\bar{\gamma}\in\mathrm{GL}(2,\Z{})\,\right]=-1\;,\quad\Phi~\xmapsto{\bar{\gamma}}~(c\bar{T}+d)^n \rho(\bar{\gamma}) \bar{\Phi}\;,\quad  T~\xmapsto{\bar{\gamma}}~\frac{a\bar{T}+b}{c\bar{T}+d}\;.
 \end{equation}
\end{itemize}
All of these symmetries are individually known from bottom-up model building, see~\cite{Feruglio:2019ybq}. In explicit top-down constructions
we find that \textit{all of these arise at the same time} in a non-trivially unified fashion~\cite{Baur:2019iai,Baur:2019kwi,Nilles:2020nnc,Nilles:2020kgo,Nilles:2020tdp,Nilles:2020gvu,Ohki:2020bpo,Nilles:2021ouu}, that we call the ``eclectic'' flavor symmetry~\cite{Nilles:2020nnc}
\begin{equation}\label{eq:eclectic}
\fcolorbox{red}{yellow}{\Large ~$G_\mathrm{\mathbf{eclectic}} ~=~ G_\mathrm{traditional} ~\cup~ G_\mathrm{modular} ~\cup~ G_\mathrm{R} ~\cup~ \CP\,.$\vspace{0.2cm}~}
\end{equation}

\section{Origin of the eclectic flavor symmetry in heterotic orbifolds}
A new insight is that in the Narain lattice formulation of compactified heterotic string theory~\cite{Narain:1985jj,Narain:1986am,GrootNibbelink:2017usl}
the complete unified eclectic flavor symmetry can unambiguously derived from the outer automorphisms~\cite{Trautner:2016ezn} of the Narain lattice space group~\cite{Baur:2019kwi, Baur:2019iai}.
These outer automorphisms contain modular transformations, including the well-known T-duality transformation and the so called mirror symmetry (permutation of different moduli) of string theory, but also symmetries of the \CP-type as well as traditional flavor symmetries and, therefore, naturally yield the unification shown in Eq.~\eqref{eq:eclectic}.
The eclectic transformations also automatically contain the previously manually derived so-called ``space-group selection rules'' \cite{Hamidi:1986vh,Dixon:1986qv,Ramos-Sanchez:2018edc} 
and non-Abelian ``traditional'' flavor symmetries~\cite{Kobayashi:2006wq}.

\section[The eclectic flavor symmetry of T2/Z3]{\boldmath The eclectic flavor symmetry of $\mathbbm T^2/\Z3$ \unboldmath}
Let us now focus on a specific example model~\cite{Baur:2021bly} in which the six extra dimensions of ten-dimensional heterotic string theory are compactified in such
a way that two of them obey the $\mathbbm T^2/\Z3$ orbifold geometry.  The discussion of this $D=2$ subspace involve a K\"ahler and complex structure modulus
$T$ and $U$, respectively, with the latter being fixed to $\langle U\rangle=\exp(\nicefrac{2\pi\I}3)=:\omega$ by the orbifold action.
The outer automorphisms of the corresponding Narain space group yield the full eclectic group of this setting, which is of order $3888$ and given by\footnote{%
Finite groups are denoted by $\mathrm{SG}\left[\cdot,\cdot\right]$ where the first number is the order of the group and the second their GAP SmallGroup ID~\cite{GAP4url}.
}~\cite{Nilles:2020tdp,Nilles:2020gvu}
\begin{equation}
  G_\mathrm{eclectic} ~=~ \Omega(2)\rtimes\Z2^\CP\,,\qquad\text{with}\quad \Omega(2)\cong \mathrm{SG}[1944,3448]\,.
\end{equation}
More specifically, $G_\mathrm{eclectic}$ contains
\begin{itemize}
\item a $\Delta(54)$ traditional flavor symmetry,
\item the $\mathrm{SL}(2,\Z{})_T$ modular symmetry of the $T$ modulus, which acts as a $\Gamma'_{3}\cong T'$ finite 
      modular symmetry on matter fields and their couplings,
\item a $\Z9^R$ discrete R symmetry as remnant of $\mathrm{SL}(2,\Z{})_U$, and 
\item a $\Z2^\CP$ \CP-like transformation.
\end{itemize}
These symmetries and their interplay are shown in table~\ref{tab:Z3FlavorGroups}. 
Twisted strings localized at the three fixed points of the $\mathbbm{T}^2/\Z3$ orbifold
form three generations of massless matter fields in the effective IR theory with 
transformations under the various symmetries summarized in table~\ref{tab:Representations}.
Explicit representation matrices of the group generators are shown in the slides of the talk and in our paper~\cite{Baur:2021bly}.
Examples for complete string theory realizations are known, see~\cite{Carballo-Perez:2016ooy,Ramos-Sanchez:2017lmj} and \cite{Baur:2021bly,Baur:2021pr}, 
and we show the derived charge assignment of the SM-like states in one particular example in table~\ref{tab:Z3xZ3configurations}.
\begin{table}[t!]
\center
\resizebox{\textwidth}{!}{
\begin{tabular}{|c|c||c|c|c|c|c|c|}
\hline
\multicolumn{2}{|c||}{nature}        & outer automorphism       & \multicolumn{5}{c|}{\multirow{2}{*}{flavor groups}} \\
\multicolumn{2}{|c||}{of symmetry}   & of Narain space group    & \multicolumn{5}{c|}{}\\
\hline
\hline
\parbox[t]{3mm}{\multirow{6}{*}{\rotatebox[origin=c]{90}{eclectic}}} &\multirow{2}{*}{modular}            & rotation $\mathrm{S}~\in~\SL{2,\Z{}}_T$ & $\Z{4}$      & \multicolumn{3}{c|}{\multirow{2}{*}{$T'$}} &\multirow{6}{*}{$\Omega(2)$}\\
&                                    & rotation $\mathrm{T}~\in~\SL{2,\Z{}}_T$ & $\Z{3}$      & \multicolumn{3}{c|}{}                      & \\
\cline{2-7}
&                                    & translation $\mathrm{A}$                & $\Z{3}$      & \multirow{2}{*}{$\Delta(27)$} & \multirow{3}{*}{$\Delta(54)$} & \multirow{4}{*}{$\Delta'(54,2,1)$} & \\
& traditional                        & translation $\mathrm{B}$                & $\Z{3}$      &                               & & & \\
\cline{3-5}
& flavor                             & rotation $\mathrm{C}=\mathrm{S}^2\in\SL{2,\Z{}}_T$      & \multicolumn{2}{c|}{$\Z{2}^R$} & & & \\
\cline{3-6}
&                                    & rotation $\mathrm{R}\in\SL{2,\Z{}}_U$   & \multicolumn{3}{c|}{$\Z{9}^R$}   & & \\
\hline
\end{tabular}
}
\caption{\label{tab:Z3FlavorGroups}
Eclectic flavor group $\Omega(2)$ for six-dimensional orbifolds that contain a 
$\mathbbm T^2/\Z{3}$ orbifold sector~\cite{Nilles:2020tdp}. 
}
\end{table}
\begin{table}[h]
\center
\resizebox{\textwidth}{!}{
\begin{tabular}{|c||c||c|c|c|c||c|c|c|c||c|}
\hline
\multirow{3}{*}{sector} &\!\!matter\!\!& \multicolumn{9}{c|}{eclectic flavor group $\Omega(2)$}\\
                        &fields        & \multicolumn{4}{c||}{modular $T'$ subgroup} & \multicolumn{4}{c||}{traditional $\Delta(54)$ subgroup} & $\Z{9}^R$ \\
                        &$\Phi_n$      & \!\!irrep $\rep{s}$\!\! & $\rho_{\rep{s}}(\mathrm{S})$ & $\rho_{\rep{s}}(\mathrm{T})$ & $n$ & \!\!irrep $\rep{r}$\!\! & $\rho_{\rep{r}}(\mathrm{A})$ & $\rho_{\rep{r}}(\mathrm{B})$ & $\rho_{\rep{r}}(\mathrm{C})$ & $R$\\
\hline
\hline
bulk      & $\Phi_{\text{\tiny 0}}$   & $\rep1$             & $1$                   & $1$                   & $0$               & $\rep1$   & $1$               & $1$                   & $+1$ & $0$         \\
          & $\Phi_{\text{\tiny $-1$}}$& $\rep1$             & $1$                   & $1$                   & $-1$              & $\rep1'$  & $1$               & $1$                   & $-1$ & $3$         \\
\hline
$\theta$  & $\Phi_{\nicefrac{-2}{3}}$ & $\rep2'\oplus\rep1$ & $\rho(\mathrm{S})$    & $\rho(\mathrm{T})$    & $\nicefrac{-2}{3}$& $\rep3_2$ & $\rho(\mathrm{A})$& $\rho(\mathrm{B})$    & $+\rho(\mathrm{C})$ & $1$\\
          & $\Phi_{\nicefrac{-5}{3}}$ & $\rep2'\oplus\rep1$ & $\rho(\mathrm{S})$    & $\rho(\mathrm{T})$    & $\nicefrac{-5}{3}$& $\rep3_1$ & $\rho(\mathrm{A})$& $\rho(\mathrm{B})$    & $-\rho(\mathrm{C})$ & $-2$\\
\hline
$\theta^2$& $\Phi_{\nicefrac{-1}{3}}$ & $\rep2''\oplus\rep1$& $(\rho(\mathrm{S}))^*$& $(\rho(\mathrm{T}))^*$& $\nicefrac{-1}{3}$& $\crep3_1$& $\rho(\mathrm{A})$& $(\rho(\mathrm{B}))^*$& $-\rho(\mathrm{C})$ & $2$\\
          & $\Phi_{\nicefrac{+2}{3}}$ & $\rep2''\oplus\rep1$& $(\rho(\mathrm{S}))^*$& $(\rho(\mathrm{T}))^*$& $\nicefrac{+2}{3}$& $\crep3_2$& $\rho(\mathrm{A})$& $(\rho(\mathrm{B}))^*$& $+\rho(\mathrm{C})$ & $5$\\
\hline
\hline
super-    & \multirow{2}{*}{$W$} & \multirow{2}{*}{$\rep1$} & \multirow{2}{*}{$1$} & \multirow{2}{*}{$1$} & \multirow{2}{*}{$-1$} & \multirow{2}{*}{$\rep1'$} & \multirow{2}{*}{$1$} & \multirow{2}{*}{$1$} & \multirow{2}{*}{$-1$} & \multirow{2}{*}{$3$}\\
\!\!potential\!\! & & & & & & & & & & \\
\hline
\end{tabular}
}
\caption{\label{tab:Representations}
$T'$, $\Delta(54)$ and $\Z9^R$ representations of massless matter fields $\Phi_n$ with modular weights $n$ 
in semi-realistic heterotic orbifold compactifications with a $\mathbbm{T}^2/\Z3$ sector~\cite{Nilles:2020kgo}. 
}
\end{table}
\begin{table}[t]
	\centering
	%\resizebox{\textwidth}{!}{ 
		\begin{tabular}{cllllllllll}
			\toprule
			& $\ell$                   & $\bar e$              & $\bar\nu$             & $q$                   & $\bar u$
			& $\bar d$              & $H_u$                 & $H_d$                 &  $\varphi_\mathrm{f}$ & $\phi^0_\mathrm{f}$\\
			\midrule
			Model A & $\Phi_{\nicefrac{-2}3}$ & $\Phi_{\nicefrac{-2}3}$ & $\Phi_{\nicefrac{-2}3}$ & $\Phi_{\nicefrac{-2}3}$ & $\Phi_{\nicefrac{-2}3}$
			& $\Phi_{\nicefrac{-2}3}$ & $\Phi_{0}$  & $\Phi_{0}$  & $\Phi_{\nicefrac{-2}3}$ & $\Phi_{0}$\\
%			B & $\Phi_{\nicefrac{-1}3}$ & $\Phi_{\nicefrac{-2}3}$ & $\Phi_{\nicefrac{-2}3}$ & $\Phi_{\nicefrac{-2}3}$ & $\Phi_{\nicefrac{-2}3}$
%			& $\Phi_{\nicefrac{-1}3}$ & $\Phi_{-1}$ & $\Phi_{0}$  & $\Phi_{\nicefrac{-2}3,-1}$ \\
%			C & $\Phi_{\nicefrac{-2}3}$ & $\Phi_{\nicefrac{-1}3}$ & $\Phi_{\nicefrac{-1}3}$ & $\Phi_{\nicefrac{-1}3}$ & $\Phi_{\nicefrac{-1}3}$
%			& $\Phi_{\nicefrac{-2}3}$ & $\Phi_{-1}$ & $\Phi_{-1}$ & $\Phi_{\nicefrac{-1}3,-1}$\\
%			D & $\Phi_{\nicefrac{-1}3}$ & $\Phi_{\nicefrac{-1}3}$ & $\Phi_{\nicefrac{\pm2}3,0}$ & $\Phi_{\nicefrac{-1}3}$ & $\Phi_{\nicefrac{-1}3}$
%			& $\Phi_{\nicefrac{-1}3}$ & $\Phi_{0}$ & $\Phi_{-1,0}$ & $\Phi_{\nicefrac{\pm2}3,-1}$\\
%			E & $\Phi_{\nicefrac{-2}3,\nicefrac{-1}3}$ & $\Phi_{\nicefrac{-2}3,0}$ & $\Phi_{0,\nicefrac{-2}3,\nicefrac{-1}3,\nicefrac{-5}3}$ & $\Phi_{-1,\nicefrac{-2}3}$ & $\Phi_{\nicefrac{-2}3}$
%			& $\Phi_{0,\nicefrac{-2}3}$ & $\Phi_{0}$ & $\Phi_{0}$ & $\Phi_{\nicefrac{-2}3,\nicefrac{-1}3,\nicefrac{-5}3,-1}$\\
			\bottomrule
		\end{tabular}
	%}
	\caption{\label{tab:Z3xZ3configurations}
		Flavor symmetry representations of MSSM quark ($q,\bar u,\bar d$), lepton ($\ell,\bar e,\bar\nu$), Higgs and flavon fields ($\varphi$,$\phi$)
		in an example of a consistent string theory configuration with a $\mathbbm T^2/\Z3$ orbifold sector.
		Following the notation of table~\ref{tab:Representations}, representations are \textit{entirely} determined by stating the respective modular weight.
}
\end{table}

Generic $\Omega(2)$ compliant super- and K\"ahler potentials have been derived in~\cite{Nilles:2020kgo} and their explicit form can be found in~\cite{Baur:2021pr}. 
For our example model A,
\begin{equation}
\begin{split}
   W ~=~  \phi^0 & \left[  \left(\phi^0_\mathrm{u}\,\varphi_\mathrm{u}\right) Y_\mathrm{u}\, H_\mathrm{u}\,\bar{u}\,q + \left(\phi^0_\mathrm{d}\,\varphi_\mathrm{e}\right) Y_\mathrm{d}\, H_\mathrm{d}\, \bar{d}\, q\, 
                     + \left(\phi^0_\mathrm{e}\,\varphi_\mathrm{e}\right) Y_\ell\, H_\mathrm{d}\, \bar{e}\, \ell\right] \\
       &              + \left(\phi^0\varphi_\nu\right) Y_\nu\, H_\mathrm{u}\,\bar\nu\, \ell  + \phi^0_\mathrm{M}\,\varphi_\mathrm{e}\,Y_\mathrm{M}\,\bar\nu\,\bar\nu \,.             
\end{split}                     
\end{equation}\enlargethispage{0.5cm}
Two important empirical observations can be made in this top-down setting: (i) While matter fields can have fractional modular weights, they always combine in such a way
that all Yukawa couplings are modular forms of \textit{integer} weight. (ii) The charge assignments under the eclectic symmetry are \textit{uniquely}
fixed in one-to-one fashion by the modular weight of a field. The latter also holds for all other known top-down constructions, see~\cite{Kikuchi:2021ogn,Baur:2020jwc,Baur:2021mtl,Almumin:2021fbk,Ishiguro:2021ccl}, 
and can be conjectured to be a general feature of TD models~\cite{Baur:2021bly}.

\section{Sources of eclectic flavor symmetry breaking}
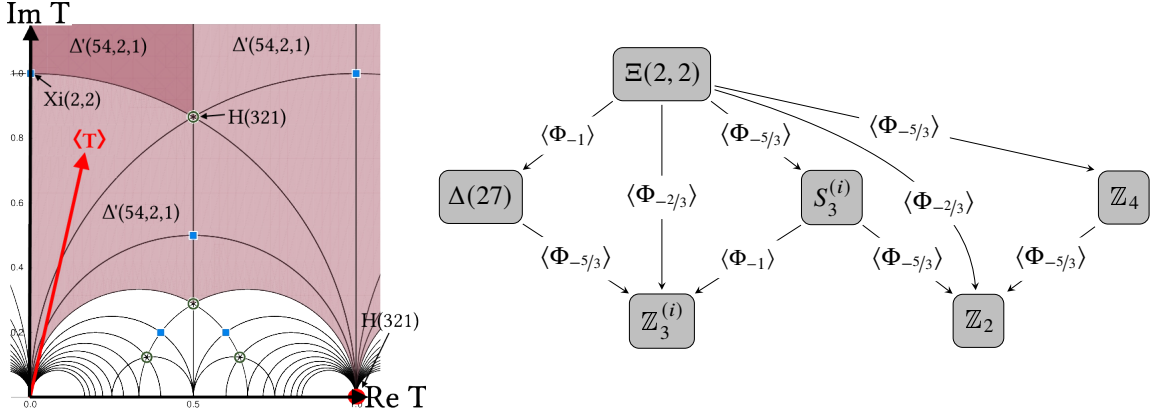
\begin{figure}[t!]
%	\centering
%\hspace{-0.5cm}
\resizebox{\textwidth}{!}{
	\includegraphics[width=0.4\linewidth]{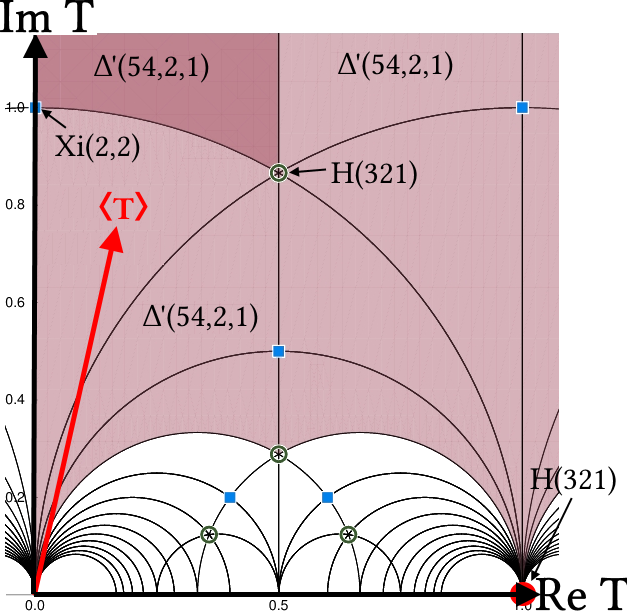}
\raisebox{0.2\height}{
	\begin{tikzpicture}[node distance=1cm and 1.3cm, rounded corners, >=stealth]
	
	\node[minimum height=22pt, draw, rectangle,fill=lightgray] (D54) {$\;\!\Xi(2,2)$ };
	\node[minimum height=22pt, draw, rectangle, below left=of D54,fill=lightgray] (D27) { $\Delta(27)$ };
	\node[minimum height=22pt, draw, rectangle, below right=of D54,fill=lightgray] (S3) { $\;\!S_3^{(i)}$ };
	\node[minimum height=22pt, draw, rectangle, below = 2.8cm of D54,fill=lightgray] (Z3) { $\;\!\Z3^{(i)}$ };
	\node[minimum height=22pt, draw, rectangle, below right= of S3,fill=lightgray] (Z2) { $\;\!\Z2$ };
	\node[minimum height=22pt, draw, rectangle, right= 3.4cm of S3,fill=lightgray] (Z4) { $\;\!\Z4$ };
	
	\draw[->, >=stealth] (D54.south west) -- node [fill=white,rectangle,midway,align=center] {\resizebox*{23pt}{!}{$\vev{\Phi_{-1}}$}} (D27.north east);
	\draw[->, >=stealth] (D54.south east) -- node [fill=white,rectangle,midway,align=center] {\resizebox*{30pt}{!}{$\vev{\Phi_{-\nicefrac53}}$}} (S3.north west); 
	\draw[->, >=stealth] (D27.south east) -- node [fill=white,rectangle,midway,align=center] {\resizebox*{30pt}{!}{$\vev{\Phi_{-\nicefrac53}}$}} (Z3.north west);
	\draw[->, >=stealth] (S3.south west) -- node [fill=white,rectangle,midway,align=center] {\resizebox*{23pt}{!}{$\vev{\Phi_{-1}}$}} (Z3.north east);
	\draw[->, >=stealth] (S3.south east) -- node [fill=white,rectangle,midway,align=center] {\resizebox*{30pt}{!}{$\vev{\Phi_{-\nicefrac53}}$}} (Z2.north west);
	\draw[->, >=stealth, shorten >=1.5pt, shorten <=1.5pt] (D54.south) -- node [fill=white,rectangle,midway,align=center] {\resizebox*{30pt}{!}{$\vev{\Phi_{-\nicefrac23}}$}} (Z3.north);
	\draw[->, shorten >=1.5pt, shorten <=1.5pt] (D54) .. controls ($(S3)+(1.4,0.65)$)  and ($(S3)+(2.,-0.45)$) .. node [fill=white,rectangle,pos=0.56,align=center] {~\resizebox*{30pt}{!}{$\vev{\Phi_{-\nicefrac23}}$}} (Z2);
	\draw[->, shorten >=1.5pt, shorten <=1.5pt] (D54) -- node [fill=white,rectangle,midway,align=center] {\resizebox*{30pt}{!}{$\vev{\Phi_{-\nicefrac53}}$}} (Z4.north west);
	\draw[->, shorten >=1.5pt, shorten <=1.5pt] (Z4.south west) -- node [fill=white,rectangle,midway,align=center] {\resizebox*{30pt}{!}{$\vev{\Phi_{-\nicefrac53}}$}} (Z2.north east);
	\end{tikzpicture}}}
	\caption{\label{fig:Xi22breaking}
		Left: Residual symmetries of the eclectic flavor symmetry $\Omega(2)$ in dependence of the modulus vev $\langle T\rangle$ 
		in the bulk of the fundamental domain and at symmetry enhanced special points.
		Right: Flavon vev induced breaking patterns of the linearly realized unified flavor symmetry $\Xi(2,2)$ at $\vev{T}=\I$.
}
\end{figure}

The eclectic flavor symmetry is broken by both, the vacuum expectation value (vev) of the modulus $\langle T\rangle$ and the vevs of flavon fields.
This is unlike in virtually all current bottom-up models where either one or the other breaking mechanism is implemented.
Note that all vevs $\langle T\rangle$ have non-trivial stabilizers in the eclectic symmetry that lead to enhancements
of the residual traditional flavor symmetry beyond what has been previously known in the literature. 
This situation is depicted in figure~\ref{fig:Xi22breaking} (left).
For a realistic phenomenology the residual traditional flavor symmetry has to be further broken by the vevs of flavon fields.
On the right of figure~\ref{fig:Xi22breaking} we exemplary show the possible breaking patterns induced by different flavon vevs 
for the particular example of a $\Xi(2,2)$ residual traditional flavor symmetry that arises for $\langle T\rangle=\I$. 
Since different residual symmetries are possible for different sectors of the theory the overall symmetry can be completely broken
even though vevs are typically stabilized at symmetry enhanced points.

\section{Possible lessons for consistent bottom-up model building}
Given this explicit example of a complete top-down model, the following empirical observations can be made 
that might be taken as guidepost for bottom-up constructions:
(i) Neither modular nor traditional flavor symmetries arise alone but they arise as mutualy overlapping parts of the full eclectic flavor symmetry, including also
\CP-type and R symmetries. (ii) Modular weights of matter fields are fractional, while modular weights of (Yukawa) couplings are integer.
(iii) Modular weights are $1:1$ ``locked'' to other flavor symmetry representations. (iv) Different sectors of the theory may have different
moduli and/or different residual symmetries allowing for what has been called ``local flavor unification''~\cite{Baur:2019kwi}.
If all these features would indeed be confirmed on other UV complete top-down constructions one may anticipate that in a modern
language, the modular flavor ``swampland'' may be much bigger than anticipated.

\section{Summary}
There are explicit models of compactified heterotic string theory that reproduce in the IR the MSSM+eclectic flavor symmetry+flavon fields.
The complete eclectic flavor symmetry here can be unambiguously computed from the outer automorphisms of the Narain space group
and it non-trivially unifies previously discussed traditional, modular, R and \CP-type flavor symmetries, see equation~\eqref{eq:eclectic}.
The eclectic flavor symmetry is broken by vacuum expectation values of the moduli and of the flavon fields. 
While residual symmetries are common, their breaking and subsequent approximate nature can help to naturally generate
hierarchies in masses and mixing matrix elements. 
Further topics to be investigated encompass the inclusion of the extra tori, the question of moduli stabilization,
the computation of the flavon potential, as well as investigating the restrictions on the K\"ahler potential, see \cite{Chen:2019ewa,Chen:2021prl},
and corresponding corrections to the phenomenological predictions.
A complete fit of this model to the observed SM flavor structure including K\"ahler corrections is underway~\cite{Baur:2021pr}.

\section*{Acknowledgements}
I would like to thank my collaborators on these subjects, Alexander Baur, Hans Peter Nilles, Saul Ramos-S\'anchez, and Patrick Vaudrevange,
also for useful comments on the manuscript.
\bibliography{Orbifold}
\bibliographystyle{JHEP}
\end{document}